\documentclass[namedreferences,hyperref,optionalrh,solaromanenum]{spr-sola}

\usepackage{graphicx}                    % For eps figures, newer & more powerfull
\usepackage{amssymb}                    % useful mathematical symbols
\usepackage{amsmath}
\usepackage{multirow}
\usepackage{lmodern}
\usepackage{color}                       % For color text: \color command
\chardef\us=`\_

\begin{document}

\begin{frontmatter}

\title{Recovery of the Solar Cycle from Maunder-like Grand Minima Episodes: A Quantification of the Necessary Polar Flux Threshold through Solar Dynamo Simulations}

%%%%%%%%%%%%%%%%%%%%%%%%%%%%%%%%%%%%%%%%%%%%%%%%%%%
%% Authors Names and Affiliations

\author[addressref=aff1,email={cs20rs018@iiserkol.ac.in}]{\inits{C.}\fnm{Chitradeep}~\snm{Saha}\orcid{0000-0002-3131-8260}}
\author[addressref={aff1,aff2,aff3},email={chandra@mps.mpg.de}]{\inits{S.}\fnm{Sanghita}~\snm{Chandra}\orcid{0000-0003-1987-0033}}
\author[addressref={aff1,aff2},corref,email={dnandi@iiserkol.ac.in}]{\inits{D.}\fnm{Dibyendu}~\snm{Nandy}\orcid{0000-0001-5205-2302}}

\address[id=aff1]{Center of Excellence in Space Sciences India, Indian Institute of Science Education and Research Kolkata, Mohanpur 741246, West Bengal, India}
\address[id=aff2]{Department of Physical Sciences, Indian Institute of Science Education and Research Kolkata, Mohanpur 741246, West Bengal, India}
\address[id=aff3]{Max-Planck-Institut f{\"u}r Sonnensystemforschung, Justus-Von-Liebig-Weg 3, 37077 G\"ottingen, Germany}

%%%%%%%%%%%%%%%%%%%%%%%%%%%%%%%%%%%%%%%%%%%%%%%%%%%
%% Runningheads
%
\runningauthor{Saha et al.}
\runningtitle{Recovering from solar grand minimum}

%%%%%%%%%%%%%%%%%%%%%%%%%%%%%%%%%%%%%%%%%%%%%%%%%%%
%%% Abstract 
\begin{abstract}
The 11-yr cycle of sunspots undergo amplitude modulation over longer timescales. As a part of this long-term modulation in solar activity, the decennial rhythm occasionally breaks, with quiescent phases with very few sunspots observed over multiple decades. These episodes are termed as solar grand minima. Observation of solar magnetic activity proxies complemented by solar dynamo simulations suggests that the large-scale solar polar fields become very weak during these minima phases with a temporary halt in the polar field reversal. Eventually, with the accumulation of sufficient polar fluxes, the polarity reversal and regular cyclic activity is thought to resume, Using multi-millennial dynamo simulations with stochastic forcing, we quantify the polar flux threshold necessary to recover global solar polarity reversal and surmount grand minima phases. We find that the duration of a grand minimum is independent of the onset rate and does not affect the recovery rate. Our results suggest a method to forecast the Sun's recovery from a grand minima phase. However, based on our approach, we could not identify specific precursors that signal entry in to a grand minima phase -- implying that predicting the onset of grand minima remains an outstanding challenge.   
\end{abstract}

%%%%%%%%%%%%%%%%%%%%%%%%%%%%%%%%%%%%%%%%%%%%%%%%%%%
% Keywords


\end{frontmatter}
%-------------------------------------------------

%%%%%%%%%%%%%%%%%%%%%%%%%%%%%%%%%%%%%%%%%%%%%%%%%%%
\section{Introduction} 
\label{sec:intro}

The Sun is a middle-aged star in the main sequence whose magnetic output fluctuates over a wide range of timescales, creating a dynamic space weather and space climate \citep{SCHRIJVER2015ADVANCESINSPACERESEARCH, PEVSTOV2023ADVSPRES}. One of the primary proxies of solar magnetic activity is the number of sunspots, that waxes and wanes in a cyclic decennial rhythm -- called the sunspot cycle \citep{HATHAWAY2015LIVREV}. Besides, there are instances when this decennial rhythm breaks, sunspots almost disappear from the solar surface, and the star enters an apparent magnetic quiescence for a prolonged period, known as the solar grand minimum \citep{USOSKIN2007AANDA, NANDY2021PEPS, Saha_Nandy_2023}. The most recent example of such a phase, as reported by various astronomers, is the Maunder Minimum, lasting nearly 70 years during 1645-1715 \citep{EDDY1976SCIENCE}. Solar grand minima episodes are manifestations of one of the extreme phases of long-term solar magnetic variability.  As evidenced in the reconstructed history of solar activity spanned over multiple millennia, the Sun has spent significant time in grand minima phases \citep{Usoskin2023LivRev}. In addition to its magnetic and particulate output, the Sun's irradiance reduces substantially during a grand minimum, which in turn influences the terrestrial climate systems \citep{SHINDELL2001SCIENCE}. Therefore, understanding the dynamics of a solar grand minimum -- its onset, development and termination -- holds significant implications including the predictability of long-term solar variability \citep{Daglis2021AG}.

Occurrence of intermittency like grand minimum in the solar magnetic activity has been studied in the past from both the perspectives of observations \citep{BEER1998SOLPHYS, USOSKIN2015AANDA, ANDRES2018NATASTRON, CARRASCO2021MNRAS, INCEOGLU2024SCIREP} as well as numerical simulations \citep{CHARBONNEAU2004APJ, MOSS2008SOLPHYS, USOSKIN2009SOLPHYS,  PASSOS2014AANDA, HAZRA2014APJ, Cameron2017ApJ, Karak2018ApJ, TRIPATHI2021MNRASL, SAHA2022MNRASL}. 
Besides a temporary cessation in the solar photospheric magnetic activity, a global change in the Sun's parity resulting from hemispheric asymmetry in the magnetic activity is a characteristic signature of solar grand minima episodes \citep{CALLEBAUT2007ADVSPRES, OLEMSKOY2013APJ, HAZRA2019MNRAS}. The parity of large-scale solar magnetic fields and the global topology of the solar corona during a grand minimum phase is predominantly determined by the Sun's polar fields  \citep{HAYAKAWA2021JSWSC, DASH2023MNRAS}. However, direct observation of the solar polar field during any of the previous solar grand minima is not available, thereby making it difficult to understand the dynamics of polar magnetic fields including their reversals during grand minima. Several independent solar dynamo simulations have reported temporary halts in the reversal of the Sun’s polar magnetic fields during deep solar grand minima \citep{MACKAY2003SOLPHYS, SAHA2022MNRASL}. Based on variations in the abundance of cosmogenic isotopes during the Maunder Minimum, \cite{MAKAROV2000JAA} concluded that the Sun's polar field reversal in the northern hemisphere might have halted during this period. Simulated polar flux time series by \cite{SAHA2022MNRASL} also demonstrated occasional halt in the hemispheric polar field reversal, resuming eventually through a gradual accumulation of magnetic flux transported by the poleward meridional circulation. 
Following a halt in polar field reversal, the process of flux accumulation continues until enough polar fields of a particular polarity are built up that can kick-start the regular sunspot cycles.

In this study, we simulate multi-millennial solar magnetic activity with intermittent grand minima episodes, and quantify the threshold of solar polar flux upon achieving which dynamo mechanism comes out of sub-criticality aiding in regular photospheric magnetic activities and denoting the recovery from these grand minima phases. We also investigate whether the onset rate, recovery rate, and duration of grand minima bear any significant underlying correlations. However, our simulations show no such relationship, implying that, in our current understanding, it is difficult to predict an impending solar grand minimum accurately. Nevertheless, once the solar activity attains a grand minimum, a regular and reliable measurement of the solar polar flux should enable us to forecast the time of corresponding recovery.

\section{Model setup} 
% \label{sec: methods}

Direct imaging of the solar interior is impossible, and existing high-latitude solar observations face significant projection issues, leaving these regions largely uncharted. Knowledge of solar interior and polar region is crucial in the context of grand minima, as photospheric polar fields  and deep-seated toroidal fields drive solar magnetic activity in the absence of sunspots \citep{NANDY2023BAAS, NANDY2023JASTP}. Magnetohydrodynamic simulation of the solar dynamo can shed light on the dynamics in the solar convection zone \citep{NANDY2002SCIENCE, CHATTERJEE2004AANDA, PASSOS2014AANDA, BHOWMIK2018NATCOM}. Here, we utilise an axisymmetric dynamo model working in the kinematic regime \citep{CHATTERJEE2004AANDA}.  The governing equations are,

% \begin{equation}
\begin{align}
    \frac{\partial B}{\partial t} + \dfrac{1}{r}\left[\dfrac{\partial}{\partial r}(rv_rB) + \dfrac{\partial}{\partial\theta}(v_\theta B)\right]  = {\eta}_t\left( \nabla^2 - \frac{1}{r^2\sin^2\theta}  \right)B \\ 
    +  r\sin\theta\left(B_p\cdot \nabla\right) \Omega   + \frac{1}{r}\frac{\partial (rB)}{\partial r}\frac{\partial {\eta}_t}{\partial r}{\notag}
\end{align}
% \end{equation}

\begin{equation}
    \frac{\partial A}{\partial t} + \frac{1}{r\sin\theta}\left[ (\mathbf{v_p} \cdot \nabla) (r\sin\theta A) \right] = {\eta}_p\left( \nabla^2 - \frac{1}{r^2\sin^2\theta}  \right)A + \alpha B
\end{equation}

\begin{figure}[hbtp]
    \centering    \includegraphics[width=0.7\linewidth]{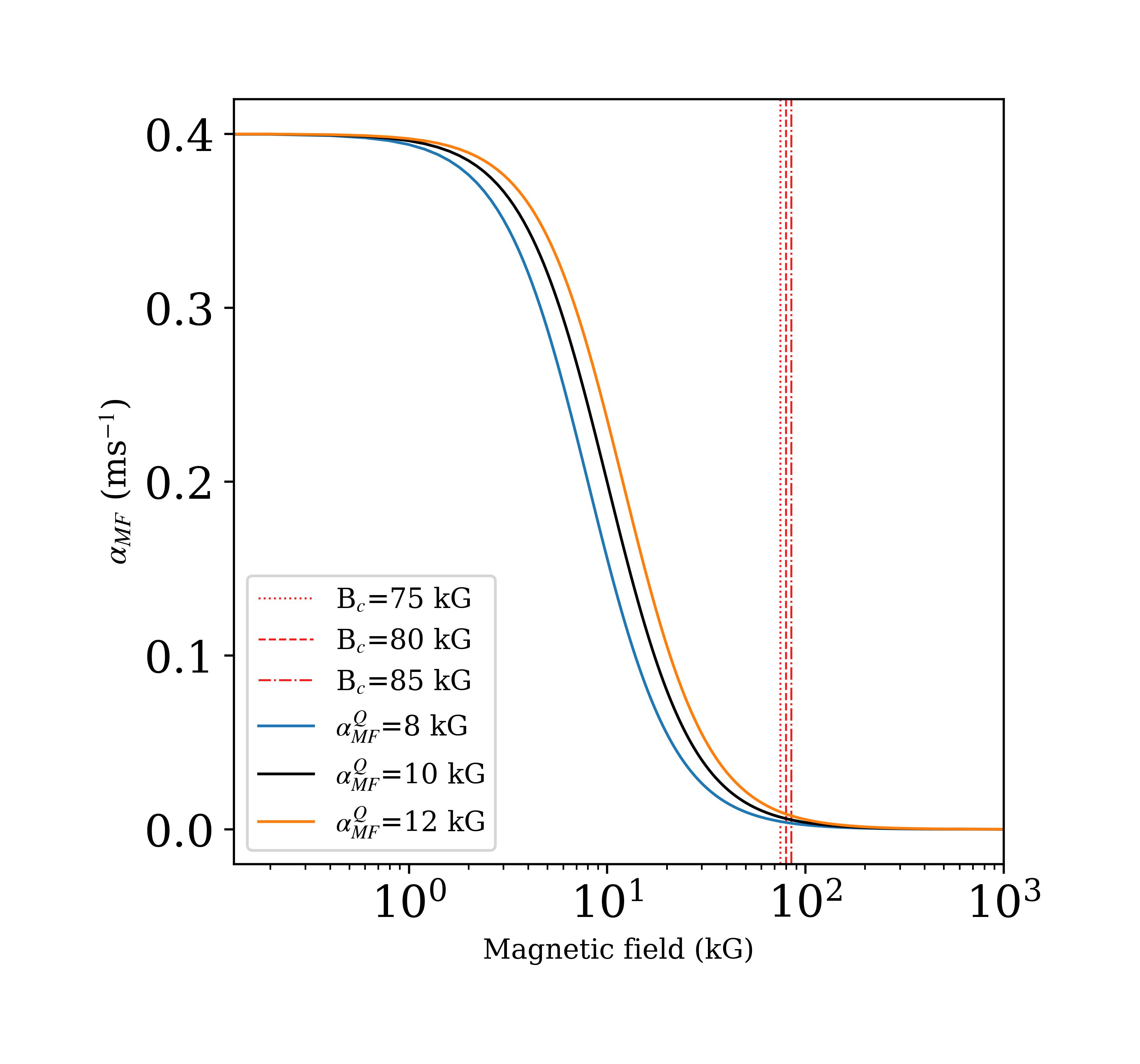}
    \caption{A visualization of the nonlinear algebraic quenching profile used for the mean-field poloidal source, $\alpha_{MF}$, in our simulations with three different choices of magnetic field magnitudes, i.e., $\alpha_{MF}^Q$ = 8, 10, and 12 kG. $B_c$ denotes the critical buoyancy threshold imposed on the deep-seated toroidal fields beyond which surface eruption events are possible in the dynamo model. $B_c$ assumes following three values, 75, 80 and 85 kG. For the primary simulation run, $\alpha_{MF}^Q$ and $B_c$ are set at 10 and 80 kG, respectively, while some other combinations of these parameters are explored in subsequent sets of simulations.}
    \label{fig:figure1}
\end{figure}

\noindent Here, $B(r, \theta, t)\mathbf{e_{\phi}}$ denotes the toroidal component of the Sun's large-scale magnetic field, and  $A(r, \theta, t) \mathbf{e_{\phi}}$ denotes the vector potential corresponding to the poloidal component.  Meridional flow is denoted by $\mathbf{v_p}$, whereas $\Omega$ represents the rotation in the solar convection zone. The symbols $\eta_p$ and $\eta_t$ denote the poloidal and toroidal diffusivities, respectively. The poloidal source term $\alpha$ is an algebraic superposition of two different sources, i.e., mean-field $\alpha_{MF}$, operating in the bulk of the convection zone, and the Babcock-Leighton $\alpha_{BL}$, operating near the surface. Spatial profiles of $\Omega$, $\mathbf{v_p}$, $\eta_p$, and  $\eta_t$ are taken from \citeauthor{CHATTERJEE2004AANDA} (\citeyear{CHATTERJEE2004AANDA}; equations 8-13, therein), whereas, \citeauthor{PASSOS2014AANDA} (\citeyear{PASSOS2014AANDA}; equations 3 \& 4) elucidates the spatial distributions of both poloidal sources along with their magnetic quenching profiles. To maintain notational consistency, we reiterate below the quenching profile of $\alpha_{MF}$ used here illustrating how the amplitude, $\alpha_{0MF}$, relates to the toroidal component, $B$, and the magnetic quenching limit, $\alpha_{MF}^Q$ (see Fig.\ref{fig:figure1}),  

\begin{equation}
   \alpha_{0MF} \sim \dfrac{\alpha_{0MF}}{1 + \left(\dfrac{B}{\alpha_{MF}^Q}\right)^2} 
\end{equation}

Long-term solar magnetic variability originates from a stochastic forcing of the nonlinear dynamo mechanism \citep{Choudhuri1992AandA, MININNI2002PRL, CAMERON2019AANDA, Saha2025ApJL}. Turbulent plasma motions in the solar convection zone stochastically perturb the buoyantly unstable toroidal flux tubes, resulting in a dispersion around their mean tilt angle. It is now widely accepted that the Sun's magnetic field cycle is driven by a Babcock-Leighton (BL) dynamo \citep{DASI2010AANDA,  CAMERON2015SCIENCE, PAL2024MNRAS, JASWAL2023MNRASL}. In the BL paradigm, the migration of diffused magnetic fluxes from anomalously tilted active regions introduces a random component in the solar polar field build-up \citep{Jiang2014ApJ, Jiang2015ApJL, NAGY2017SolPhys, Karak2017ApJ, PAL2023APJ, PKumar2024MNRAS}.  Observations of titled bipolar magnetic regions suggest that the BL process may be subject to Gaussian noise \citep{Jiang2014ApJ, Karak2017ApJ, PAL2023APJ}. However, we note that complex interactions among dynamical variables can alter the statistical nature of the stochastic noise -- shifting, for example, from uniform white noise to autocorrelated colored noise -- as described in \cite{Saha2025ApJL}, which employed a similar model setup. Following Occam's razor philosophy, we introduce stochastic fluctuations in the form of uniform white noise in both poloidal sources in our simulations  independently in both hemispheres.

Numerical values of various parameters including the level of fluctuations in the poloidal sources are chosen in order to faithfully reproduce the occurrence statistics of grand minima gleaned from multi-millennial scale reconstruction of solar magnetic activity \citep{USOSKIN2007AANDA, Usoskin2023LivRev}. Table \ref{table:table2} in the Appendix contain a list of key parameter values used in our simulations.

In order to further explore causal correlations between relevant physical parameters, we perform additional sets of simulations by varying the critical buoyancy threshold, B$_c$,  and the magnetic quenching limit, $\alpha_{MF}^Q$, in the mean-field poloidal source as summarized in Fig.\ref{fig:figure1}. We restrict all our simulations to a near-critical dynamo number regime -- the likely regime of operation of the solar dynamo currently \citep{Metcalfe2016ApJL, vanSaders2016Nature, REINHOLD2020SCIENCE, TRIPATHI2021MNRASL, Wavhal2025}.

\section{Results \& Discussions} 
\label{sec: results}

\subsection{Dynamics of hemispheric unipolar phases during simulated grand minima}

\begin{figure}
    \centering
    \includegraphics[width=\linewidth]{ 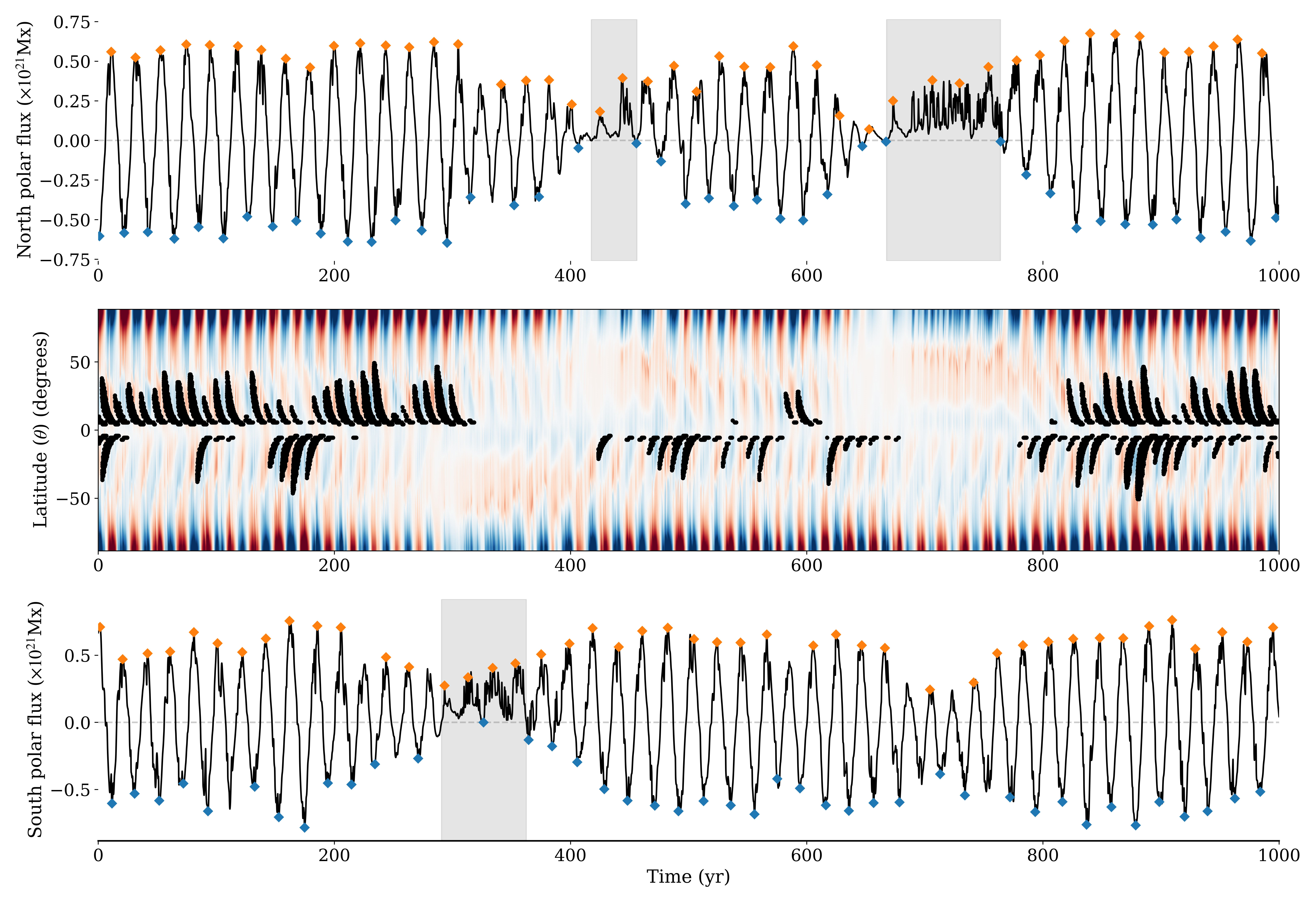}
    \caption{An overview of the temporal dynamics of simulated solar polar fields over a millennium. The top and bottom panels show the evolution of northern and southern hemispheric polar fields, with grand minima-like intermittent phases. Episodes with no polarity reversal are shaded in grey. Red and blue markers track the cycle amplitudes for positive and negative polarities, respectively. The middle panel depicts the surface magnetic butterfly diagram for the corresponding time frame, with a color-coded signed radial field in the background and sunspot eruption proxies (black) in the foreground. The color map is saturated to a value of $\pm$150 G. In the illustrated 1000-yr simulation there are two distinct hemispheric grand minima in each hemisphere.
}
    \label{fig:Figure2}
\end{figure}

We calculate the flux of poloidal magnetic fields penetrating the solar photosphere in the latitudinal extent of 70$^{\circ}$ to 87$^{\circ}$ over both hemispheres and ascribe this as the hemispheric polar flux in our analysis. Flux of toroidal magnetic field is calculated near the base of the convection zone (0.68R$_\odot$-0.73R$_\odot$) around active latitudes i.e., 3$^{\circ}$-45$^{\circ}$ in both hemispheres. Extended episodes with little to no surface eruption proxies in any hemisphere for at least three consecutive solar cycles are considered to be grand minima episodes in our simulations. As demonstrated earlier by \cite{SAHA2022MNRASL}, these simulations reveal a specific sequence of events at the onset and during grand minima i.e., a decline in polar flux, followed by a reduction in toroidal flux amplitudes. As the toroidal field weakens, sunspots disappear. The polar flux continues to drop until it reaches a minimal value, after which the polar field ceases to reverse due to inadequate surface field strength. From now on, we refer the phases with no polarity reversal as `unipolar' phases. Notably, such unipolar phases in our simulations are largely devoid of any surface eruption (see Fig.\ref{fig:Figure2}). Nevertheless, the large scale plasma motions along with turbulent diffusion work in tandem to drive a weak internal dynamo and dredge up deep seated toroidal fields towards the surface. The mean-field poloidal source takes over the toroidal fields and inducts poloidal fields that eventually migrate towards the polar caps. Polar fields start reversing once enough flux from a particular polarity builds up thereby revitalizing the regular dynamo cycles. Fig.\ref{fig:Figure2} elucidates this sequence for a couple of hemispheric grand minima episodes. 

At this point we investigate if there is any threshold value for the polar flux below which it triggers a grand minimum like dormancy in the solar magnetic activity and vice-versa. We track the signed hemispheric polar flux amplitude right before and after the unipolar phases throughout the simulated time series spanning over multiple millennia. Our analysis finds no specific amplitude threshold, falling below which the polar field reversal halts. This suggests that the onset of unipolar phases is less likely a gradual and systematic phenomenon. Notably, unipolar phases in our simulations occur well into the grand minima, i.e., after the sunspot proxies disappear from the solar surface. In that way, the existence of a threshold for the onset of unipolar phases is not so important from a predictive point of view.

\begin{figure}
    \centering
    \includegraphics[width=\linewidth]{ 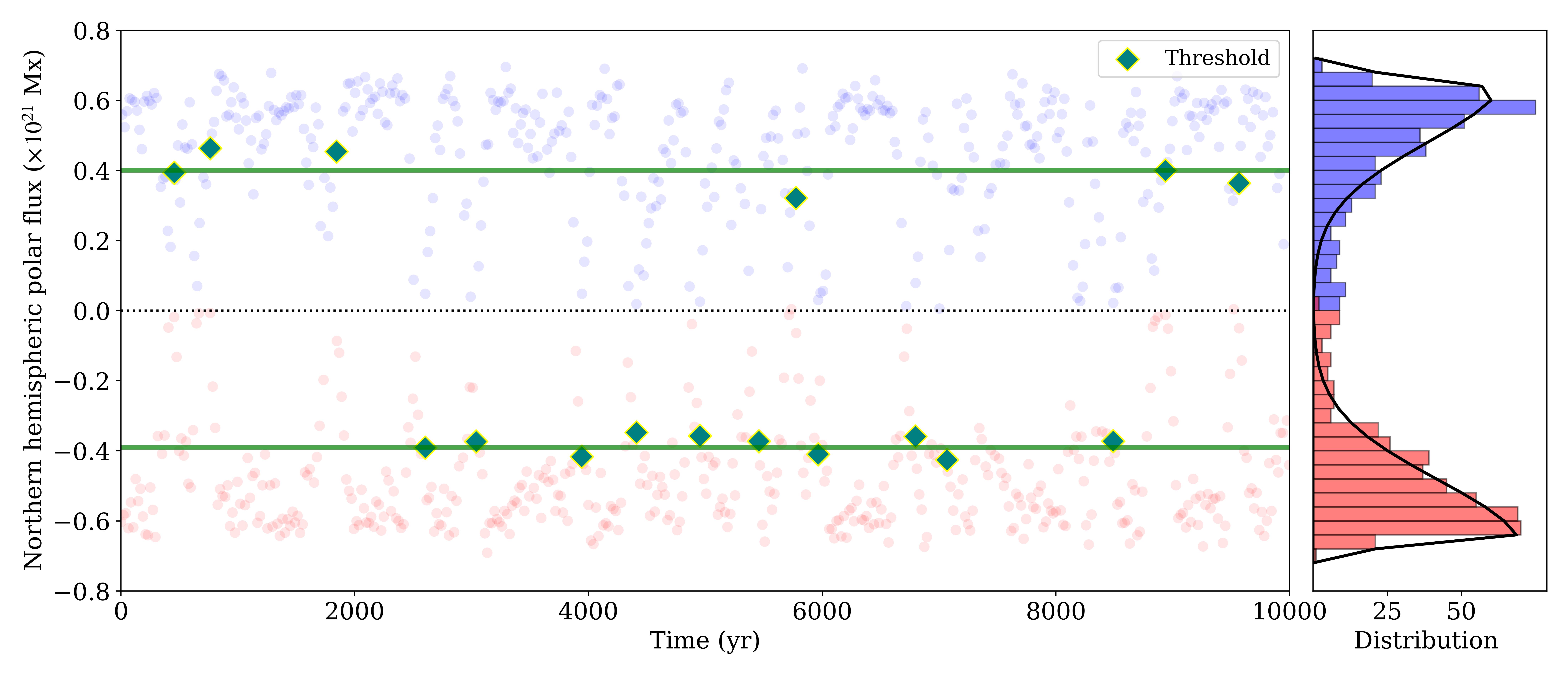}
    \includegraphics[width=\linewidth]{ 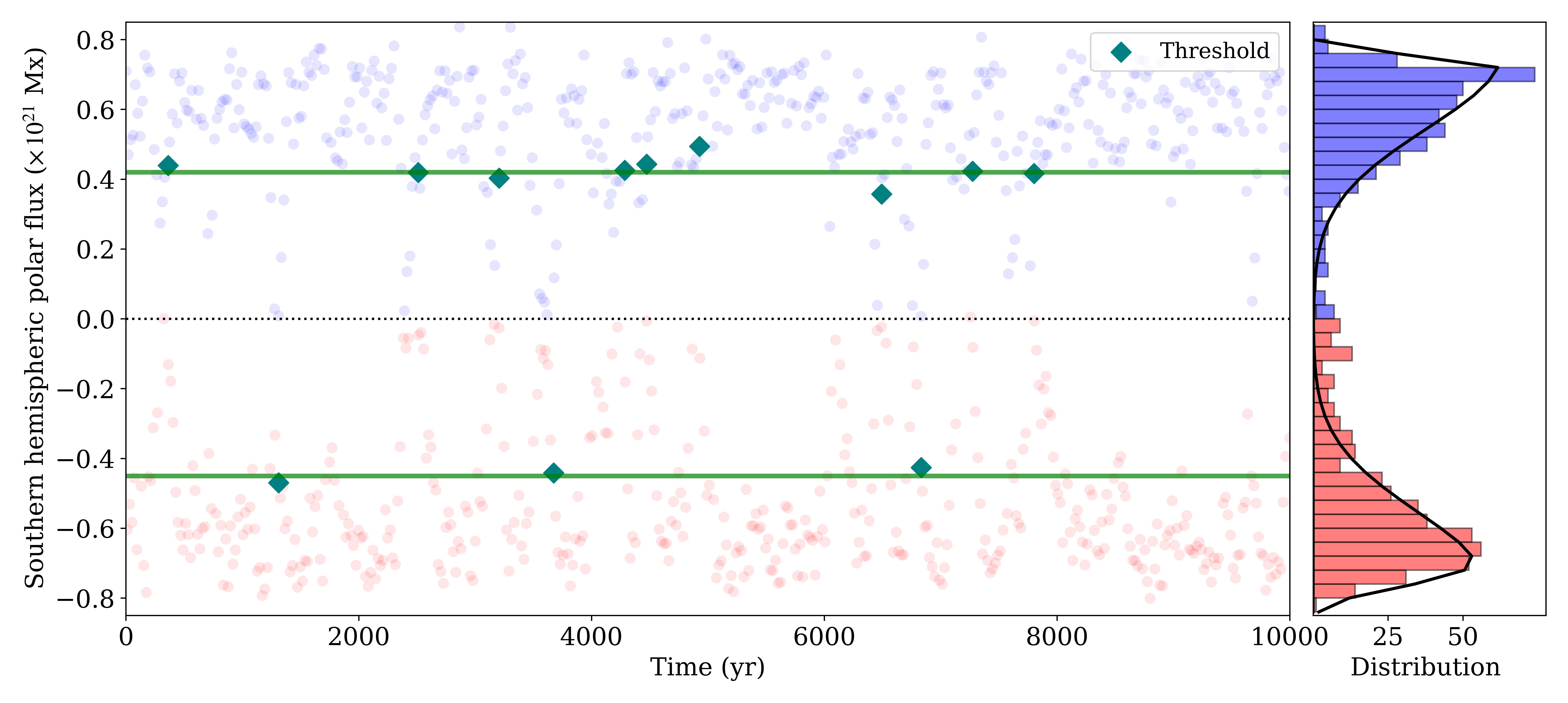}
    \caption{Distribution of simulated polar flux amplitudes (blue: positive; red: negative) over ten millennia, for northern (top panel) and southern (bottom panel) solar hemispheres. Corresponding histograms depict the distribution of polar flux amplitudes over the global timescale. The polar fields remain predominantly in the regular activity phase,  therefore the histograms are fitted with Gaussian curves with non-zero skewness. Green diamond markers denote those amplitude values associated with the termination of unipolar phases. The clustering of these amplitudes around a particular mean value (marked in green straight line) indicates a systematic threshold for the polar flux that exists to facilitate recovery from grand minima episodes. Notably, not all data points with lower polar flux in our simulations belong to grand minima -- even during the regular magnetic activity period, discrete occurrence of weak polar flux cycle weaker than the recovery threshold is possible.}
    \label{fig:fighem}
\end{figure}

Intriguingly, we discover that the signed hemispheric polar flux amplitude lies within a certain narrow range of values right before the recovery from the unipolar phase. This clustering of amplitude values has potentially significant implication for predicting the recovery from a solar grand minimum. The top and bottom panels in Fig.\ref{fig:fighem} demonstrate this clustering for north and south polar fields, respectively. We present an estimate of the threshold around which the recovery flux amplitudes are clustered by taking into account all the polar flux cycle amplitudes for a given simulation run (see the histograms in Fig.\ref{fig:fighem}). The recovery threshold can be quantified with respect to the mode associated with the skewed-Gaussian distribution of signed hemispheric polar flux amplitudes. This modal value determines the typical amplitude scale corresponding to the most recurrent value of the polar field. For the simulation shown in Figure \ref{fig:Figure2}, the threshold lies within a range of $63\pm4\%$ of the modal polar flux amplitude. This holds true for both polarities across both hemispheres. We note that even during the regular magnetic activity period, discrete occurrence of weak polar flux cycle weaker than the recovery threshold is possible.

\begin{figure}
    \centering
    \includegraphics[width=0.75\linewidth]{ 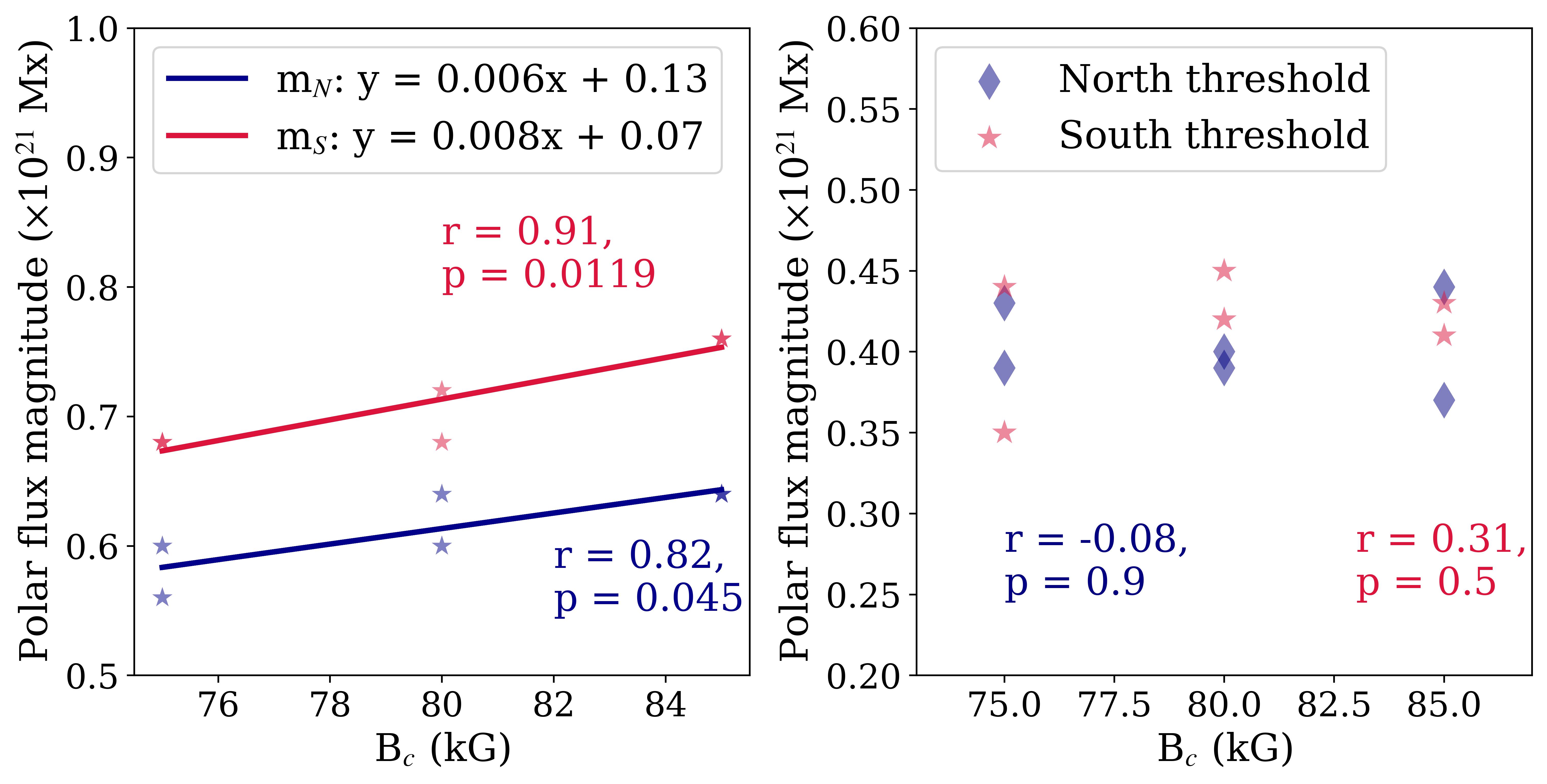}
    \caption{Modal values of hemispheric polar flux are linearly correlated to the critical buoyancy threshold, B$_c$ (left panel; m$_N$ for north, in blue and m$_S$ for south, in red). Whereas, the recovery threshold polar flux is independent of B$_c$ (right panel; threshold for northern hemisphere in blue diamonds, and for southern hemisphere in red asterisks). The correlation coefficient and the parameter of statistical significance are denoted by $r$ and $p$, respectively.}
    \label{fig:Figure5}
\end{figure}

\begin{figure}
    \centering
    \includegraphics[width=\linewidth]{ 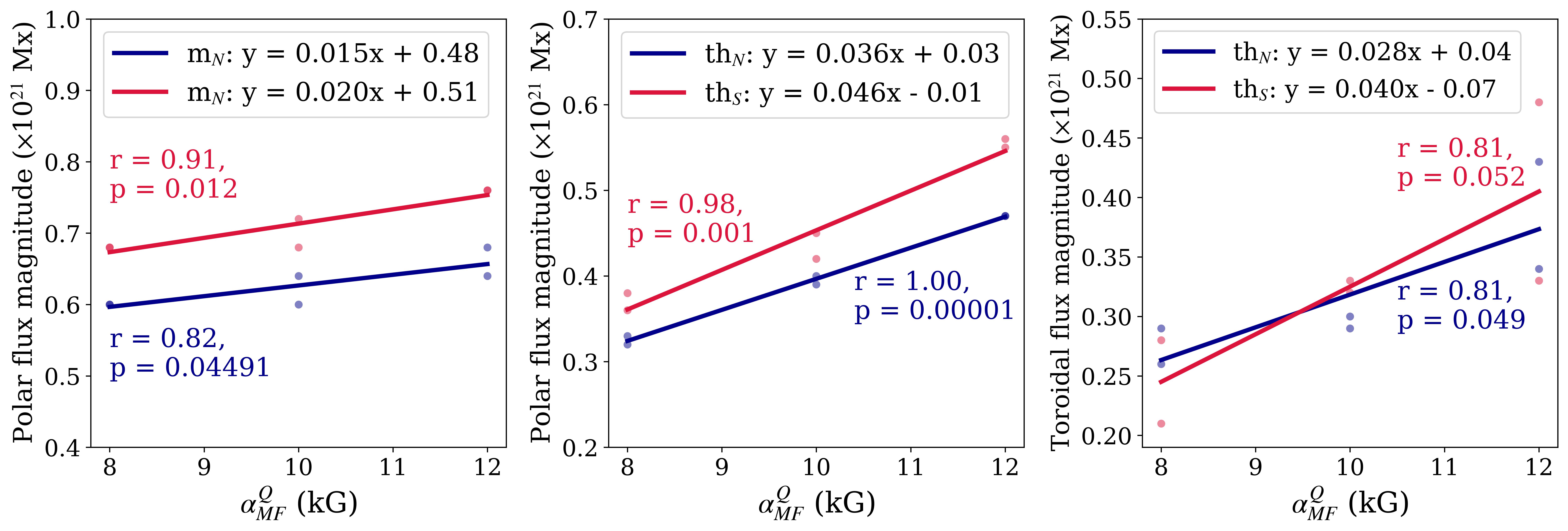}
    \caption{Evidence of statistically significant correlation between the (a) modal values of hemispheric polar flux (left panel; m$_N$ for north, in blue and m$_S$ for south, in red), (b) recovery threshold values of hemispheric polar flux (middle panel; th$_N$ for north, in blue and th$_S$ for south, in red), (c) corresponding hemispheric toroidal flux (right panel; th$_N$ for north, in blue and th$_S$ for south, in red).  In each case, the corresponding linear regression is presented. The correlation coefficient and the parameter of statistical significance are denoted by $r$ and $p$, respectively.}
    \label{fig:Figure6}
\end{figure}

\subsection{Exploring causal correlations}

Besides invoking stochastic forcings on the poloidal sources, our simulations also incorporate a couple of nonlinear mechanisms that can potentially impact estimating this polar flux threshold value. In our model, there are nonlinear magnetic field quenching at play for both mean-field poloidal source, $\alpha_{MF}$, and the Babcock-Leighton poloidal source, $\alpha_{BL}$.  The deep seated $\alpha_{MF}$ inducts poloidal fields effectively from weaker toroidal fields ($<$10 kG, for our primary run), whereas,  $\alpha_{BL}$ acts on toroidal fields lying within the range of 1-500 kG to contribute to generating poloidal fields near the surface. Very weak toroidal flux tubes get shredded by turbulent plasma motions in the convection zone, while very strong ones are immune to the Coriolis force, causing them to emerge with no considerable tilt \citep{Weber2011ApJ, Fan2021LivRev} and consequently making them ineffective for poloidal field induction \citep{PASSOS2014AANDA}. It is important to note that, $\alpha_{BL}$ plays little to no role well in to grand minima as these phases lack surface eruption events. Therefore, $\alpha_{MF}$ is the predominant poloidal source governing the magnetic field dynamics during our simulated grand minima episodes.

Since the dynamo operates slightly above the critical dynamo number regime in our simulations, the imposition of a critical magnetic buoyancy threshold, $B_c$ implies an additional source of nonlinearity. $B_c$  regulates the emergence of deep-seated toroidal fields onto the photosphere \citep{NANDY2001APJ, Nandy2002ASpSc}. We use B$_C$ = 80 kG as the critical buoyancy limit for our primary run.

Now, we explore how $B_c$ and $\alpha_{MF}^Q$ impact the polar field dynamics during the unipolar phase in a grand minimum. From a value of 80 kG for $B_c$, we change it to 75 kG and 85 kG, respectively, while keeping $\alpha_{MF}^Q$ =10 kG unchanged. For another set of simulations we fix $B_c$ at 80 kG and vary $\alpha_{MF}^Q$ to 8 and 12 kG, respectively. Fig.\ref{fig:figure1} summarizes this parameter space that we explore in this study. We note that the distribution of signed hemispheric polar flux amplitude and the associated modal values are dependent on the choice of $B_c$ and $\alpha_{MF}^Q$. However, the recovery polar flux in each of these simulations is still clustered around a mean value depending on the parameters -- this further bolsters the existence of a systematic and well-defined recovery polar flux threshold. The numerical values are tabulated in Table \ref{table:table1} in the Appendix.

A statistically significant linear correlation between the modal hemispheric polar flux amplitude and $B_c$ emerges from the former set of simulations. However, the corresponding recovery thresholds are independent of $B_c$ (see Fig.\ref{fig:Figure5}). This is physically justified as because during a unipolar phases in a grand minimum, the dynamo mechanism mostly inducts weak toroidal fields in the convection zone that lie below the critical buoyancy threshold. This makes the buoyancy threshold redundant during a grand minimum. 

The latter set of simulations reveals strong linear correlations between $\alpha_{MF}^Q$ and three key parameters: the modal hemispheric polar flux, the recovery polar flux threshold, and the corresponding toroidal flux (see Fig.\ref{fig:Figure6}). These correlations underscores the important role of mean-field poloidal source during grand minima. An increase in $\alpha_{MF}^Q$ diminishes the quenching; stronger toroidal fields gets more easily inducted by the mean-field poloidal source. As a result, the dynamo mechanism starts producing stronger poloidal and toroidal magnetic fluxes, thereby shifting the modal value of polar flux amplitudes to greater amplitudes. Consequently, the recovery polar flux threshold also goes up, implying a harder and slower recovery from an ongoing grand minimum.

\begin{figure}
    \centering
    \includegraphics[width=\linewidth]{ 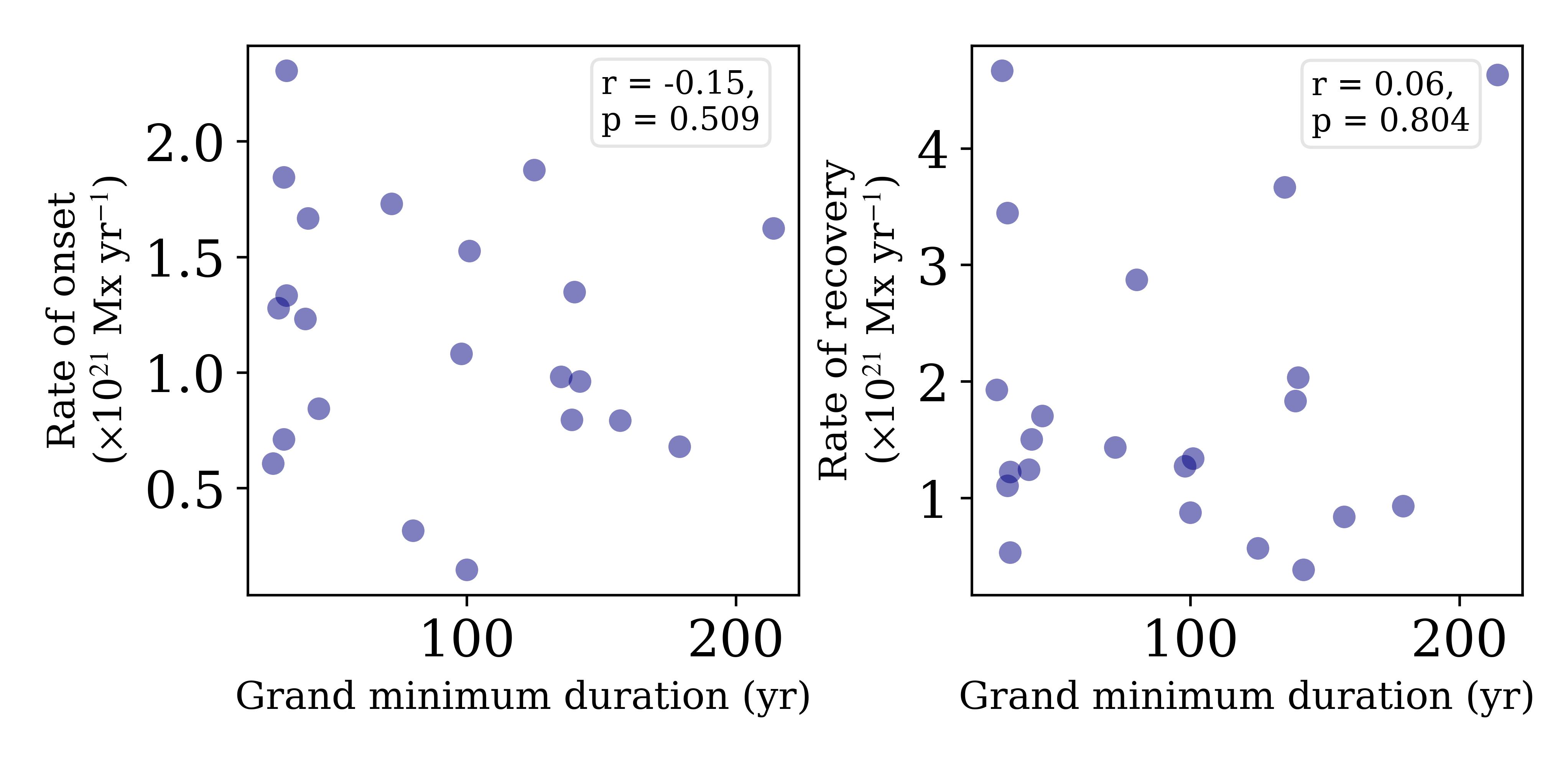}
    \caption{Linear correlation analyses between the duration of a given simulated grand minimum episode with the corresponding rate of onset (left panel) and the rate of recovery (right panel), respectively. The correlation coefficient and the parameter of statistical significance are denoted by $r$ and $p$, respectively. In both cases, no evidence of statistically significant correlation is found.  }
    \label{fig:Figure7}
\end{figure}

\subsection{Rates of onset and recovery vs. duration of grand minima}
From a predictive perspective, it is of utmost interest to know whether the duration of a grand minimum episode is related to the rate at which the amplitudes of consecutive sunspot cycle amplitudes decline at the onset of that grand minimum, or, how the amplitudes build up at its termination. Motivated by this, we consider three consecutive sunspot cycles during the onset and recovery phases of all the 22 simulated grand minima and examine their enveloping decay and rise rates, respectively. The results are presented in Fig.\ref{fig:Figure7}, clearly demonstrating the absence of any statistically significant correlation between the aforementioned quantities. Recently, \cite{INCEOGLU2024SCIREP} reported correlations between the rate and duration of onset and termination phases. Our results are independent and complementary to their findings; we conclude that the duration of the grand minimum in its entirety does not depend on the rate of onset, nor does it influence the rate of recovery from the grand minimum in our simulations.

\section{Concluding Remarks} 
% \label{sec: discussion}

Using stochastically forced multi-millennial timescale simulations of solar dynamo which exhibits intermittent grand-minima like episodes, we have investigated the nature and dynamics of solar polar fields during grand minima. Based on our simulated results, we have noted that there may be occasional halts in the polar field reversal -- signifying unipolar phases for the polar field cycle -- once the solar dynamo slips into a grand minimum. This is in agreement with observational results which suggest temporary halts in reversal \cite{MAKAROV2000JAA}. The reversal in polarity recurs only when a sufficient amount of polar flux accumulates at the solar polar caps, mediated by fluctuations in the polar field sources and large-scale plasma motions in the solar convection zone. The termination of such unipolar phases heralds the recovery of the sunspot cycle from a grand solar minimum in our simulations. \cite{SAHA2022MNRASL} report that although polarity reversals were observed to cease in their simulations, weak amplitude toroidal field cycles persist during grand minima phases. Independent work by \cite{Karak2018ApJ} suggests downward turbulent pumping also keeps weaker cycles going in the solar interior. Our work elucidates how stochastic fluctuations may recover normal solar activity levels from such weak cycles. 

We discover the existence of a (statistically inferred) threshold polar flux amplitude, upon achieving which, the regular polar flux reversal kick-starts and aid in recovery from grand minima phases. This quantification has important implications for forecasting the onset of recovery from an ongoing grand minimum episode. 

By establishing causal relationships, We demonstrate that the mean-field poloidal source $\alpha_{MF}$ plays a crucial role in determining the recovery if the polar flux threshold as noted in earlier studies \citep{HAZRA2014APJ, PASSOS2014AANDA}.

We also report that the rate at which the solar cycles enter into a magnetically quiescent grand minimum and the rate at which they recover do not correlate to the duration for which the Sun resides in that grand minimum in our simulations.
 
We also note that, in spite of an exhaustive analysis, we failed to identify any precursors or signatures in the polar flux threshold that signals entry into a grand minima phase. We leave this for future studies, emphasizing that predicting the onset of solar grand minima remains an outstanding challenge.

%%%%%%%%%%%%%%%%%%%%%%%%%%%%%%%%%%%%%%%%%%%%%%%%%%%%%%%%%%%%%%%%%%%%%%%%%%%
% Appendix

\begin{figure}
    \centering    \includegraphics[width=\linewidth]{ 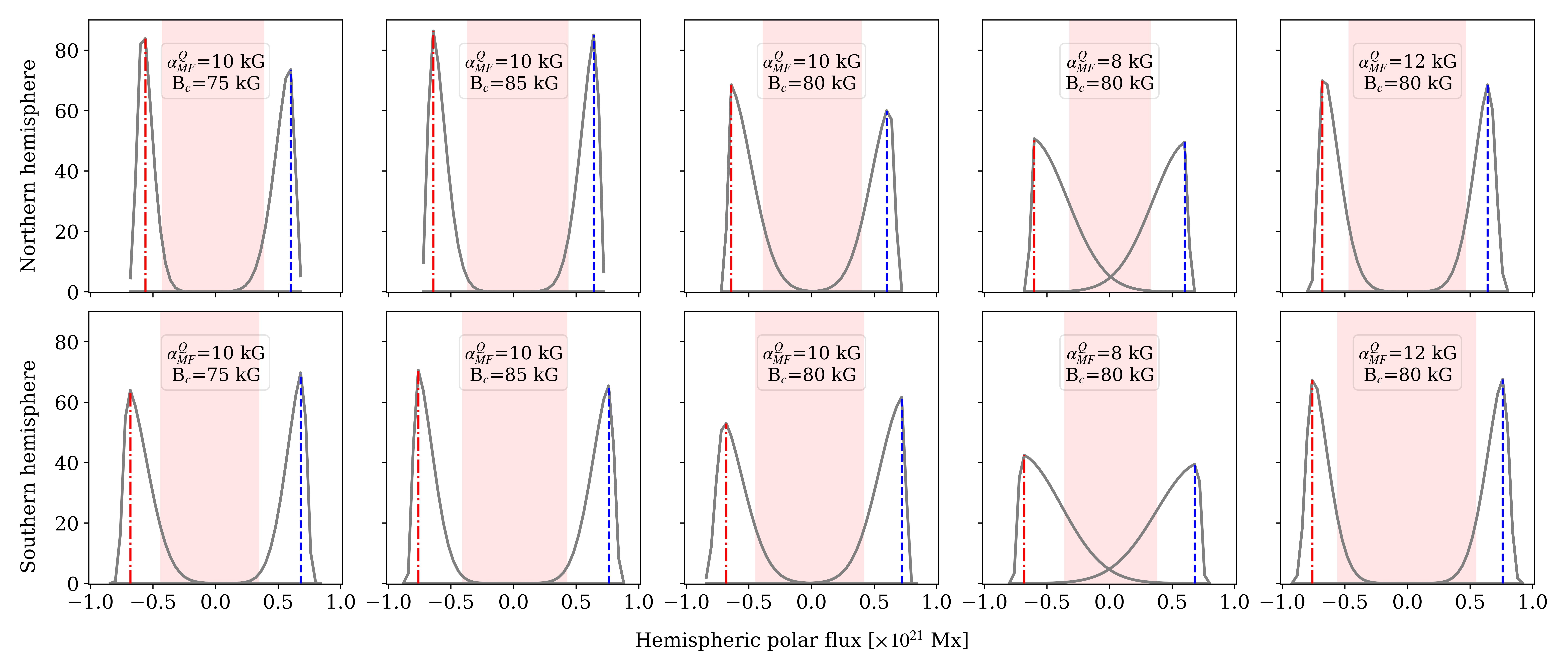}
    \caption{A visual summary of signed hemispheric polar flux distribution for different combinations of critical buoyancy threshold, $B_c$, and mean-field magnetic quenching, $\alpha_{MF}^Q$. Blue and red dashed lines mark the positive and negative modal values of the corresponding skewed Gaussian distribution, respectively. The region between positive and negative recovery polar flux  thresholds is shaded in pink.}
    \label{fig:Figure8}
\end{figure}

\appendix   
As described in previous sections, the underlying distribution of signed hemispheric polar flux amplitude, including their modal value and recovery threshold is sensitive to changes in simulation parameters like critical buoyancy threshold, $B_c$, and mean-field magnetic quenching, $\alpha_{MF}^Q$.
Fig.\ref{fig:Figure8} summarizes the above dependence, while Table \ref{table:table1} presents the corresponding values of these physical quantities.

 \begin{table*}[ht]
    \centering    
    \caption{A summary of hemispheric polar flux amplitudes and their corresponding modal and threshold values for different critical buoyancy limits, B$_C$, used in our simulations.}
    \begin{tabular}{c c c c c c c c c c}
         \hline \hline
          \multirow{1}{*}{$\alpha_{MF}^{Q}$}&\multirow{1}{*}{B$_C$}&\multirow{1}{*}{Solar}& \multicolumn{2}{c}{Polar flux (+ve)} & & \multicolumn{2}{c}{Polar flux (-ve)} \\
          \cline{4-5} \cline{7-8}
          (kG)&(kG)& Pole  & Mode  & Threshold  &  &  Mode  & Threshold   \\
          &  &  & $(10^{21}$Mx)  & $(10^{21}$Mx) &   &  $(10^{21}$Mx)  & $(10^{21}$Mx) \\
          \hline
          \multirow{2}{*}{10} &  \multirow{2}{*}{75} & North & 0.60 & 0.39  &    & -0.56 & -0.43  \\
                              &                      & South & 0.68 & 0.35  &    & -0.68 & -0.44  \\
                                \hline
          \multirow{2}{*}{10} &  \multirow{2}{*}{80}  & North & 0.60 & 0.40  &    & -0.64 & -0.39  \\
                              &                       & South & 0.72 & 0.42  &    & -0.68 & -0.45   \\
                                \hline          
          \multirow{2}{*}{10} &   \multirow{2}{*}{85}  & North & 0.64 & 0.44  &    & -0.64 & -0.37  \\
                              &                        & South & 0.76 & 0.43  &    & -0.76 & -0.41  \\
                             \hline
          \multirow{2}{*}{8}  &   \multirow{2}{*}{80}  & North & 0.60 & 0.33  &    & -0.60 & -0.32  \\
                              &                        & South & 0.68 & 0.38  &    & -0.68 & -0.36  \\
                             \hline
          \multirow{2}{*}{12}  &   \multirow{2}{*}{80} & North & 0.64 & 0.47  &    & -0.68 & -0.47  \\
                              &                        & South & 0.76 & 0.55  &    & -0.76 & -0.56  \\
          \hline
          \hline
    \end{tabular}
    \label{table:table1}
\end{table*}

In Table \ref{table:table2}, we list the values of some key parameters used in our simulations.

\begin{table*}[ht]
    \centering    
    \caption{A summary of key parameter values used in our simulations.}
    \begin{tabular}{c c c c c c c c c c}
         \hline \hline          
         Peak   & Toroidal  & Poloidal & Mean-field  & BL & Mean-field  &  BL  &  \\
         meridional  & diffusivity  & diffusivity & alpha    & alpha  & alpha  &  alpha  & \\
         speed (m/s) & (cm$^2$/s) & (cm$^2$/s) & (m/s)   & (m/s) & fluctuation  &  fluctuation  & \\
         \hline
         29  & $4\times10^{10}$  & $2.6\times10^{12}$ & 0.4  & 27 & 100$\%$  &  150$\%$  &  \\
          \hline     
          \hline
    \end{tabular}
    \label{table:table2}
\end{table*}

\newpage
%%%%%%%%%%%%%%%%%%%%%%%%%%%%%%%%%%%%%%%%%%%%%%%%%%%%%%%%%%%%%%%%%%%%%%%%%%%
%% Acknowledgements
%
\begin{acks}
CESSI is funded by IISER Kolkata, Ministry of Education, Government of India.
CS acknowledges a CSIR PhD fellowship through grant no. 09/921(0334) /2020-EMR-I. SC acknowledges the INSPIRE scholarship from the Department of Science and Technology, Government of India during her tenure as an Integrated BS-MS student at IISER Kolkata. The authors acknowledge the reviewers for useful suggestions. 
\end{acks}

%% Available additional data environments:
%% required: authorcontribution, fundinginformation, dataavailability
%% optional: materialsavailability, codeavailability
\begin{authorcontribution}
The simulations were carried out by CS in consultation with DN. CS and SC rendered the plots. All three authors contributed to interpretation of the results. CS led the writing of the paper, SC and DN assisted in revising and editing the draft. 
\end{authorcontribution}
\begin{dataavailability}
Long-term simulation data may be made available based on reasonable requests from the scientific community.
\end{dataavailability}
%
% \begin{ethics}
% \begin{conflict}
%
% \end{conflict}
% \end{ethics}

% \newpage 
%%% %%%%%%%%%%%%%%%%%%%%%%%%%%%%%%%%%%%%%%%%%%%%%%%%%%%%%%%%%%%
%% Bibliography
%
% Using BibTeX
%
\bibliographystyle{spr-mp-sola}
\bibliography{references}  

\end{document}